\documentclass[%
 aip,
 amsmath,amssymb,
 reprint,%
]{revtex4-1}

\usepackage{graphicx}
\usepackage{subfigure}
\usepackage{dcolumn}
\usepackage{bm}

\usepackage[utf8]{inputenc}
\usepackage[T1]{fontenc}
\usepackage{mathptmx}
\usepackage{amssymb}
\usepackage{epstopdf}

\usepackage[breaklinks=true,
colorlinks=true,
linkcolor=blue,
citecolor=blue,
urlcolor=blue]{hyperref}

\begin{document}

\preprint{AIP/123-QED}

\title{Effects of frequency-modulated pump on stimulated Brillouin scattering in inhomogeneous plasmas}

\author{Y. Chen}
\affiliation{School of Electrical and Information Engineering, Anhui University of Science and Technology, Huainan,Anhui 232001, China}

\author{C. Y. Zheng}
\affiliation{Institute of Applied Physics and Computational Mathematics, Beijing, 100094, China}
\affiliation{\mbox{HEDPS, Center for Applied Physics and Technology, Peking University, Beijing 100871, China}}
\affiliation{\mbox{Collaborative Innovation Center of IFSA (CICIFSA), Shanghai Jiao Tong University, Shanghai 200240, China}}

\author{Z. J. Liu}
\affiliation{Institute of Applied Physics and Computational Mathematics, Beijing, 100094, China}
\affiliation{\mbox{HEDPS, Center for Applied Physics and Technology, Peking University, Beijing 100871, China}}

\author{L. H. Cao}
\affiliation{Institute of Applied Physics and Computational Mathematics, Beijing, 100094, China}
\affiliation{\mbox{HEDPS, Center for Applied Physics and Technology, Peking University, Beijing 100871, China}}
\affiliation{\mbox{Collaborative Innovation Center of IFSA (CICIFSA), Shanghai Jiao Tong University, Shanghai 200240, China}}

\author{C. Z. Xiao } 
  \email{xiaocz@hnu.edu.cn}
  \affiliation{\mbox{Collaborative Innovation Center of IFSA (CICIFSA), Shanghai Jiao Tong University, Shanghai 200240, China}}
  \affiliation{Key Laboratory for Micro-/Nano-Optoelectronic Devices of Ministry of Education, School of Physics and Electronics, Hunan University, Changsha, 410082, China}

\date{\today}
  \begin{abstract}

 The effects of the frequency-modulated pump on stimulated Brillouin scattering (SBS) in a flowing plasma are investigated by theoretical analysis, three-wave simulations, and kinetic simulations. The resonance point of SBS oscillates in a certain spatial region with time when frequency modulations are applied. There exists a certain frequency modulation that makes the velocity of resonant points close to the group velocity of seed laser, which increases the SBS reflectivity. And the SBS could be suppressed by frequency modulation with larger bandwidth. In the kinetic simulations, the effects of the frequency-modulated pump on the reflectivity agree with our theoretical predications. The muti-location autoresonance is also observed in narrow bandwidth frequency modulation case, which can also increase the SBS reflectivity. Our work provides a method for selecting laser bandwidth to inhibition SBS in inhomogeneous plasmas.

  \end{abstract}


  \maketitle

\section{Introduction}\label{introduction}

 Stimulated Brillouin scattering (SBS) is a common instability in inertial confinement fusion (ICF)\cite{ICF1,ICFn,ICF2,ICF3,D_D,I_D,H_D,S_I}, it contributes most of the reflectivity when the pump laser passes through the plasma region of the hohlraum\cite{sbs1,sbs2,sbs3,sbs4,sbs5}, because SBS is not sensitive to the inhomogeneity of plasma density, it can occur below $n_{c}$, where $n_{c}$ is the critical density of pump laser\cite{kruer,Nicholson}. For many years, some technics about lasers are applied to mitigate parametric instabilities, such as continuous phase plates (CPPs)\cite{cpps}, spectral dispersion (SSD)\cite{ssd}, and laser pulses with spike trains of uneven duration and delay (STUD)\cite{stud}. Recently, Liu $et$ $al.$ used a multicolor alternating-polarization laser to reduce the reflectivity of SBS\cite{liu_ps1,liu_ps}.

 The broadband pump laser is a frequently-used way to suppress laser-plasma instabilities. There are two ways of making broadband lasers: muti-frequency beamlets \cite{qing,luo_1} and frequency modulation\cite{luo_2}. Zhao $et$ $al.$ have used these two ways to suppress the linear growth rate of stimulated Raman scattering (SRS)\cite{zhao_1,zhao_2}, and in their frequency modulation model, the bandwidth needs to be larger than the growth rates of SRS. In early experimental works\cite{VI1,VI2}, Arkhipenko et.al. found that the frequency-modulated pump can decrease the threshold of parametric instability and suppress the backscattering of instabilities. The influence of frequency-modulated pump on plasma parametric decay instability(PDI) is  theoretically and experimentally studied, the pump frequency modulation has both enhancement and suppression effects on PDI in inhomogeneous plasma\cite{VI3}. Ma $et$ $al.$ used the sunlight-like laser to nearly eliminate the growth of stimulated Raman side scattering (SRSS) in two-dimensional particle in cell (PIC) simulations\cite{ma}. Bates $et$ $al.$ use the pump laser with $0.6\%$ bandwidth to suppress the cross-beam energy transfer in direct-drive ICF\cite{cbet}. Recently, Wen $et$ $al.$ studied the movement of resonant points of SRS in an inhomogeneous plasma when frequency modulation of the pump laser is present\cite{wen}. Inspired by this work, we conjecture that the frequency modulation pump will also make the movement of SBS resonant points in a flowing plasma, and the frequency modulation may be an effective way to suppress the SBS reflectivity.

 Flowing plasmas are commonly observed in direct-drive\cite{D_D} and shock ignition\cite{HaoL1,Klimo1,Klimo2}. The flowing velocity of plasma is supersonic at low density region, $i.e.$ the flowing velocity is larger than the ion acoustic velocity. And the flowing velocity may influence the Landau damping of ion acoustic waves\cite{wangqing2}. Thus, the influence of flowing velocity should be taken into consideration when investigating SBS in ICF. In this paper, we consider the frequency modulation effects on SBS in a plasma with nonuniform flowing velocity. First, we theoretically study the movement of resonant points when frequency modulation is applied, and find that the effective interaction length achieves maximum when $\eta$ (define later) is close to $1$,  and the velocity of resonant points is close to the group velocity of the seed laser, where $\eta = \frac{\beta\omega_{m}^{2}L_{V}}{\omega_{00}C_{s}}$, $\beta$ represents the modulation frequency depth, $\omega_{m}$ is the modulated frequency, $L_{V}$ is the flow scale length, $\omega_{00}$ is the frequency of unmodulated pump laser, and $C_{s}$ is the ion acoustic velocity. Through three-wave simulations, we find that the SBS reflectivity also gets maximum when $\eta$ is near to $1$, and SBS reflectivity is reduced when $\eta > 2.3$.

  Then the Vlasov-Maxwell code called Vlama is used to study the nonlinear effects induced by frequency-modulated pump, we find that the autoresonance not only occurs in no bandwidth case\cite{wangqing2,wangqing1}, but also occurs in  $\eta \sim 1$ case, which is called the muti-locations autoresonance. The SBS reflectivity also increases as $\eta$ increases to $1$, and then decreases with $\eta$, which agrees well with three-wave simulations. At last, one-dimensional PIC simulations are used to verify our conclusions, the SBS reflectivity in different cases agrees with Vlasov simulations, and the movement of resonant points for particle trapping is consistent with our theoretical predictions.

 Our results can provide some guidance for the experiments. If one uses frequency-modulated pump lasers to suppress the instabilities in flowing plasma, the $\eta \sim 1$ type modulations should be avoided, and $\eta > 2.3$ type frequency modulations are preferable choices.

 This paper is structured in the following ways. Firstly, in Sec.~\ref{theoretical model}, we describe the three-wave coupling equations analysis of SBS in an flowing plasma. Secondly, Vlasov simulations and PIC simulations to verify our theoretical results are performed in Sec.~\ref{vlasov Simulation model}. At last, the conclusion  and discussion about the sinusoidal density modulation are shown in  Sec.~\ref{conclusion}.

\section{Theoretical analysis: the effects of frequency modulation on SBS}\label{theoretical model}

\subsection{ The movement of resonant point of SBS}\label{twe}

SBS is a laser plasma instability that usually observed in ICF, and the dominant process of SBS is back scattering. The resonant exaction of SBS required the phase matching conditions of pump laser, seed laser and ion acoustic wave, $\vec{k}_{0}=\vec{k}_{1}+\vec{k}_{2}$, $\omega_{00}=\omega_{1}+\omega_{2}$. $\vec{k}_{0}$, $\vec{k}_{1}$ and $\vec{k}_{2}$ are the wavenumber of three waves, $\omega_{00}$,$\omega_{1}$ are the frequency of pump laser and seed laser, $\omega_{2}=k_{2}(C_{s}+V(x))$ is the frequency of ion acoustic wave (IAW), where $C_{s} = \sqrt{(ZT_{e}+3T_{i})/m_{i}}$ is the ion sound speed, $Z$ is the charge state of ions, $T_e$ and $T_i$ are the temperature of electrons and ions, $m_{i}$ is the mass of ions, $V(x)$ is the flowing speed of plasmas.

Based on early works\cite{wangqing2,wangqing1,rosenbluth1,rosenbluth2}, there is only one resonant point in an inhomogeneous plasma though the SBS is not sensitive to the non-uniformity of plasma density. the phase mismatch will be larger when the plasma has non-uniform flowing velocity.

In an inhomogeneous plasma with flow, wavenumber detuning rate is obtained by
  \begin{equation}\label{kaba}
   \kappa'  \approx -\frac{1+M}{2}\frac{n_{e}(x_{0})}{1-n_{e}(x_{0})}\frac{k_{2}C_{s}}{V_{2}L_{n}}+\frac{k_{2}C_{s}}{V_{2}L_{V}},
  \end{equation}the first term comes from the plasma non-uniformity and the second term is from the plasma flow, $M = V(x)/C_{s}$ is the mach number, $L_{n} = n_{e}/(\partial n_{e}/\partial x)$ is the density scale length and $L_{V} = C_{s}/(\partial V/\partial x)$ is flow scale length. Based on early work\cite{rosenbluth1,rosenbluth2}, the gain of seed laser is obtained by
 \begin{equation}\label{gain}
   G_{r} = \pi \gamma_{0}^{2}/|\kappa'V_{1}V_{2}|,
  \end{equation} $\gamma_{0}$ is the growth rate of SBS at the resonant point, $V_{0}$ and $V_{1}$ are the group velocity of pump laser and seed laser, $V_{2} = (C_{s}+V(x))/c$ is the group velocity of IAW in flowing plasmas. When there is no frequency modulation of pump laser, $i.e.$ $\delta\omega = 0$,
 the resonant point is fixed at $x_{0}$ and the interaction length is obtained from Wentzel-Kramers-Brillouin(WKB) analysis\cite{rosenbluth1},

  \begin{figure}[htbp]
    \begin{center}
      \includegraphics[width=0.5\textwidth,clip,angle=0]{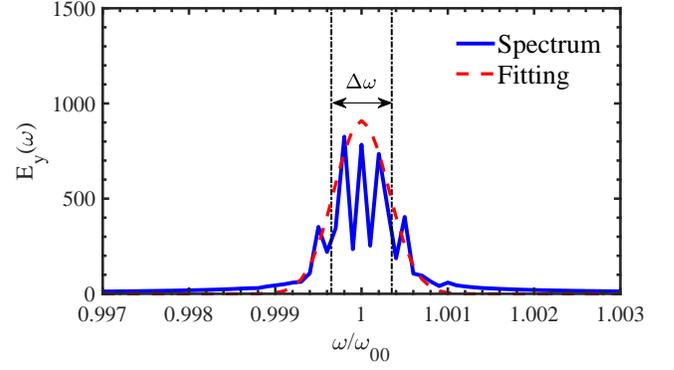}\vspace{-10pt}
      \caption{\label{theory_pump} The blue line is the frequency spectrum of pump laser when $\beta = \pi/2$ and $\omega_{m} = 2.3556\times 10^{-4}\omega_{00}$. The red dashed line represents the fitting line of spectrum by Gaussian function. And the bandwidth of pump laser is approximate to the full width at half maximum (FWHM) of Gaussian function.}
    \end{center}
  \end{figure}

\begin{figure}[htbp]
    \begin{center}
      \includegraphics[width=0.5\textwidth,clip,angle=0]{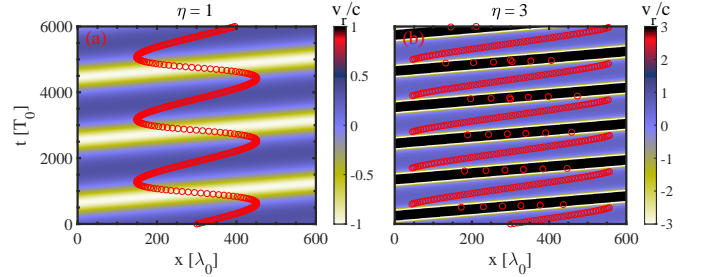}\vspace{-10pt}
      \caption{\label{theory_fig} Red circles are the resonant points oscillate with time and the colormap represents the velocities of resonant points.(a) The resonant points and corresponding velocity for $\eta = 1$ case. (b) The resonant points and corresponding velocity for $\eta = 3$ case. The definition of $\eta$ can be found in Eq.~(\ref{resonant-point-v4})  }
    \end{center}
  \end{figure}
\begin{equation}\label{gain1}
   L_{int} = 2\sqrt{\frac{2\gamma_{0}^{2}}{|V_{1}V_{2}|}}.
  \end{equation}

  The influence of frequency-modulated pump laser on stimulated Raman scattering in inhomogeneous plasma is studied by Wen et.al. they found that the movement of resonant point induced by the broadband pump beam will lead to the change of the SRS reflectivity. However, the effects of broadband pump beam on SBS in a flowing inhomogeneous plasma remain elusive. The frequency-modulated laser has a simple profile $ E_{y}(t) = a_{0}cos[\omega_{00}t+\beta sin(\omega_{m}t)]$, Fig.~\ref{theory_pump} is the spectrum of pump laser when $\beta = \pi/2$,$\omega_{m} = 2.3556\times 10^{-4}\omega_{00}$.  The pump laser has a instant frequency change $\delta\omega = \omega_{m}\beta cos(\omega_{m}t-\frac{\omega_{m}}{c}x)$, the resonant point will no longer be fixed.
  the new resonant point is called $x_{r}$, and it will oscillate near $x_{0}$, where $\omega_{m}$ is the modulation frequency, and $\beta = \pi/2$.

  Considering an inhomogeneous plasma with non-uniform flowing velocity, $V(x)=C_{s}(7/4-3x/2x_{end})$, $x_{end}$ is the total length of plasma and $c$ is the light speed in vacuum. The frequency of seed laser is equal to $\omega_{00}$, the flowing
  velocity of plasma cancels with ion sound speed at resonant point, x0 when the pump laser is monochromatic light. As for the frequency-modulated pump case, we use phase match condition, $ \omega_{0}(x,t)=\omega_{1}+\omega_{2}(x)$, to find the equation for resonant point,
  \begin{equation}\label{resonant-point}
    \omega_{00}-\omega_{1}-\frac{2C_{s}x_{r}}{cL_{V}}+\omega_{m}\beta sin(\omega_{m}t-\frac{\omega_{m}}{c}x_{r})=0,
  \end{equation}

  The velocity of resonant point is obtained by applying time derivative of Eq.~(\ref{resonant-point}),
  \begin{equation}\label{resonant-point-v}
    v_{r} = \frac{\cos(\omega_{m}t-\frac{\omega_{m}}{c}x_{r})}{\frac{2\omega_{00}C_{s}}{\beta \omega_{m}^{2}L_{V}}+\cos(\omega_{m}t-\frac{\omega_{m}}{c}x_{r})}c.
  \end{equation} If $\beta\omega_{m}\rightarrow0$, the resonant point will not not move, which is same to the situation without frequency modulation. As we know, the velocity of seed laser is close to $-c$, the SBS gain will be increased when the velocity of resonant point is close to the velocity of seed laser\cite{wen}, which is
   \begin{equation}\label{resonant-point-v2}
     \frac{\cos(\omega_{m}t-\frac{\omega_{m}}{c}x_{r})}{\frac{2\omega_{00}C_{s}}{\beta \omega_{m}^{2}L_{V}}+\cos(\omega_{m}t-\frac{\omega_{m}}{c}x_{r})} = -1,
   \end{equation}
   in the time window when the velocity of resonant point is negative, $\frac{2\omega_{00}C_{s}}{\beta \omega_{m}^{2}L_{V}} = -2\cos(\omega_{m}t-\frac{\omega_{m}}{c}x_{r})$, where $\omega_{m}t-\frac{\omega_{m}}{c}x_{r} = n\pi$, $n$ is an odd integer to make sure that the location of maximum velocity of resonant point is close to $x_{0}$. Thus, the maximum SBS gain is achieved when
   \begin{equation}\label{resonant-point-v3}
    \frac{\omega_{00}}{\beta\omega_{m}^{2}}=\frac{L_{V}}{C_{s}}.
   \end{equation}

   Here we define a indicator $\eta$ to quantify the SBS gain affected by pump's frequency modulation,
    \begin{equation}\label{resonant-point-v4}
     \eta = \frac{\beta\omega_{m}^{2}L_{V}}{\omega_{00}C_{s}},
    \end{equation} when $\eta = 1$, the SBS gain becomes maximum. In the small bandwidth limit with $\eta\leq 1$, the effective interaction length increases with frequency bandwidth,
    \begin{equation}\label{resonant-point-v5}
     L_{eff} = L_{int}+\Delta L,
    \end{equation} $\Delta L$ is is proportional to $ \triangle\omega $ and $ L_{V}$, where $\triangle\omega = 2\beta\omega_{m}$ is the bandwidth of pump laser related to the frequency modulation\cite{zhao_1,wen}. As shown in Fig.~\ref{theory_fig}(a), when $\eta =1$, the red circles show the resonant points in space and time, and the color map shows the velocity of resonant points, which is obtained by  Eq.~(\ref{resonant-point-v}). In the negative velocity time window, the maximum velocity of resonant points equal to $-c$, which is favor to the growth of SBS. However, when $\eta >1$, $i.e.$ frequency modulation is large, the negative velocity time window becomes short, the growth of SBS will be suppressed. As shown in Fig.~\ref{theory_fig}(b), where $\eta = 3$, the negative velocity time window is much shorter than Fig.~\ref{theory_fig}(a).  Next, we will prove our predications by kinetic simulations.

\subsection{ The growth rate and reflectivity of SBS with frequency-modulated pump   }\label{twe2}

\begin{figure}[htbp]
    \begin{center}
      \includegraphics[width=0.47\textwidth,clip,angle=0]{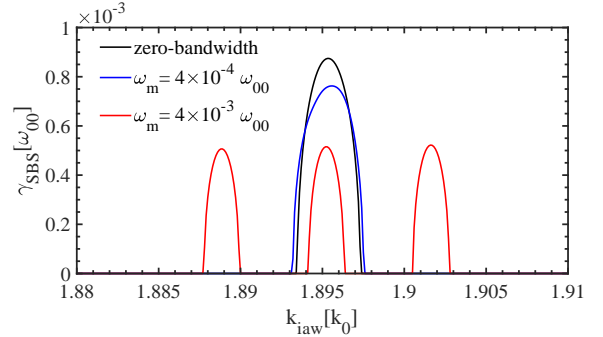}\vspace{-10pt}
      \caption{\label{growth_rate} The growth rates of SBS obtained by solving Eq.~(\ref{sbs_ds}). Black line is the growth rate of SBS with monochromatic pump, blue line is the growth rate of SBS with frequency-modulated pump when $4\times 10^{-4} \omega_{00}$, and red line is the growth rate of SBS with frequency-modulated pump when $4\times 10^{-3} \omega_{00}$. }
    \end{center}
  \end{figure}

As shown in Fig.~\ref{theory_pump}, the frequency-modulated laser has three important modes: $a_{01}(\omega_{00}-\omega_{m})$, $a_{02}(\omega_{00})$ and $a_{03}(\omega_{00}+\omega_{m})$, where $a_{01}$, $a_{02}$ and $a_{03}$ are the corresponding normalized amplitude. We also observe that three mode have almost the same amplitude. So, the normalized amplitudes of these mode can be approximate to $\frac{\sqrt{3}}{3}a_{0}$, where $a_{0}$ is the normalized amplitude of monochromatic light. From the coupling equations of SBS with muti-frequency lasers:
 \begin{equation}\label{sbs_1}
     (\frac{\partial^{2}}{\partial t^{2}}-c^{2}\nabla^{2}+\omega_{pe}^{2})\tilde{A}=-4 \pi Z ec^{2}\tilde{n}_{i} (a_{01}+a_{02}+a_{03}),
 \end{equation}

\begin{equation}\label{sbs_2}
     (\frac{\partial^{2}}{\partial t^{2}}-C_{s}^{2}\nabla^{2})\tilde{n}_{i}=\frac{n_{0}e}{m_{i}} \nabla^{2}[(a_{1}+a_{2}+a_{3})\cdot\tilde{A}],
 \end{equation} the dispersion relation of SBS can be obtained,

 \begin{equation}\label{sbs_ds}
   \omega^{2}-k^{2}C_{s}^{2} = \sum^{3}_{j=1}\frac{k^{2}a_{0j}^{2}}{4}\omega_{pi}^{2}\left[\frac{1}{D_{+,j}}+\frac{1}{D_{-,j}}\right],
 \end{equation} where $D_{\pm,j}(k,\omega)=(\omega\pm\omega_{j})^{2}-(k\pm k_{j})^{2}c^{2}-\omega_{pe}^{2}$, Solving  Eq.~(\ref{sbs_ds}), the growth rate of SBS with muti-frequency lasers can be obtained. Zhao et.al. \cite{zhao_2} found that the muti-frequency lasers will couple to one single SBS process when
 \begin{equation}\label{sbs_wm}
  \omega_{m}/\omega_{00} \leq 2a_{0}\frac{\omega_{pi}}{\omega_{00}}\sqrt{\frac{\omega_{00}}{k_2C_{s}}},
 \end{equation} the multi-frequency lasers share the wavenumber region of IAW, where $\omega_{pi} = \sqrt{Z\frac{m_{e}}{m_{i}}}\omega_{pe}$ is the ion frequency, and $\omega_{pe}$ is the electron frequency of plasma.

\begin{figure}[htbp]
    \begin{center}
      \includegraphics[width=0.5\textwidth,clip,angle=0]{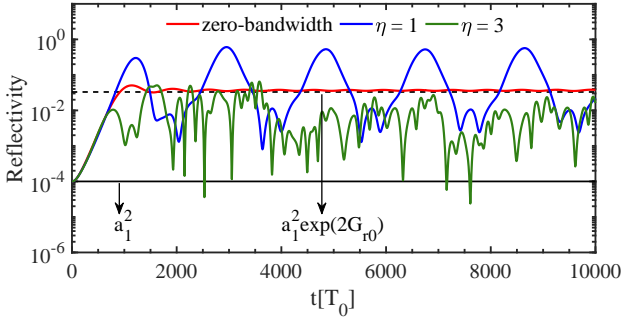}\vspace{-10pt}
      \caption{\label{tr_ref} Three-wave simulation results. The red line is the time dependence of SBS reflectivity with monochromatic pump laser. The blue line is the time dependence of SBS reflectivity for $\eta = 1$ case. The green line is the time dependence of SBS reflectivity for $\eta = 3$ case. The black dashed line is obtained by the Rosenbluth theory. }
    \end{center}
  \end{figure}

 As shown the blue line in Fig.~\ref{growth_rate}, when $\omega_{m}=4\times 10^{-4} \omega_{00}$, SBS is in the coupling regime based on Eq.~(\ref{sbs_wm}), the resonant region of SBS is only one peak, which is similar to the monochromatic light case (black line). However, when $\omega_{m}=4\times 10^{-3} \omega_{00}$, SBS is in the decoupling regime, the resonant region of SBS changes to three independent peaks, which means three independent SBS processes. In this paper, the $\omega_{m}$ is in the level of $10^{-4}\omega_{00}$ which in the coupling regime, thus, we can approximately use the three-wave model to describe the SBS process when the frequency modulation of pump is small.

 From the coupling equations of SBS, we obtain the three-wave model of SBS with frequency-modulated pump,

 \begin{equation}\label{sbs_3_wave}
    (\partial_{t}+V_{0}\partial_{x}+\nu_{0})a_{0}=-\frac{i}{4}Z\delta n_{i}a_{1}e^{-i\varphi_{1}},
\end{equation}

\begin{equation}\label{sbs_3_wave2}
    (\partial_{t}+V_{1}\partial_{x}+\nu_{1})a_{1}=-\frac{i\omega_{00}}{4\omega_{1}}Z\delta n_{i}^{\ast}a_{0}e^{i\varphi_{1}},
\end{equation}

\begin{equation}\label{sbs_3_wave3}
    (\partial_{t}+V_{2}\partial_{x}+\nu_{2}+iV_{2}\kappa'(x-x_{0}))\delta n_{i}=-\frac{-i4\gamma_{0}^{2}c^{2}}{Z\omega_{00}^{2}v_{osc}^{2}}a_{0}a_{1}^{\ast}e^{i\varphi_{1}},
\end{equation} where $a_{0}$,$a_{1}$ and $\delta n_{i}$ are the normalized amplitudes of  pump laser, seed laser and ion density perturbations of IAW, respectively. $\varphi_{1} = \int_{0}^{t}\delta \omega(x,t') dt' $ is the phase mismatch induced by frequency modulation of pump. The detail derivations are shown in Appendix.~\ref{AFluid theory}.

In a flowing inhomogeneous plasma, the flowing velocity and plasma density should fulfill the mass conservation law, $\partial(n_{e}V)/\partial x = 0$. Let us assume a linear flowing velocity in low density plasma, $V(x) = C_{s}(7/4-3x/2x_{end})$, where $x_{end} = 600 \lambda_0$ is the length of simulation box, $\lambda_0 = 351\rm{nm}$ is the wavelength of pump laser. If we choose the resonant density $n_{e0} = 0.1n_{c}$, the plasma density profile can be deduced by mass conservation law, $n_{e}(x) = n_{e0}C_{s}/V(x)$. It should be noted that the resonant plasma density locates at the center of the simulation box, and the flowing velocity of plasma at that point is equals to $-C_{s}$. The wavelength of seed laser is equal to $351\rm{nm}$ because of the zero frequency of IAW at resonant point $x_{0}$. The intensities of pump laser and seed laser are $3\times 10^{15} \rm{W/cm^{2}}$ and $3\times 10^{11} \rm{W/cm^{2}}$, respectively.

\begin{figure}[htbp]
    \begin{center}
      \includegraphics[width=0.5\textwidth,clip,angle=0]{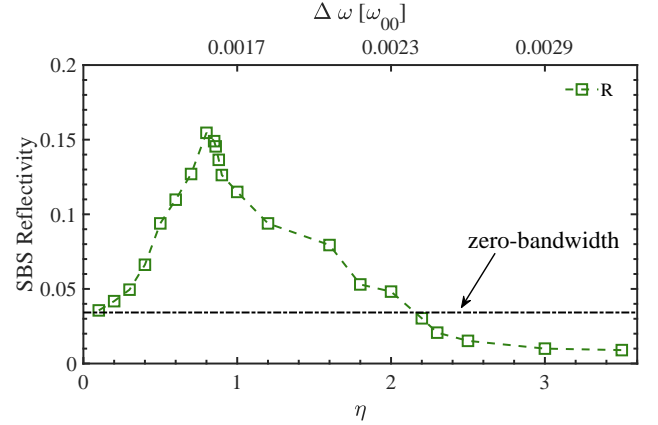}\vspace{-10pt}
      \caption{\label{ref_yita} Three-wave simulation results. The green squares are the average SBS reflectivity for different frequency modulations. The black dashed line represents the average reflectivity without frequency modulation. The top axis is the corresponding bandwidth of pump laser for each $\eta$.   }
    \end{center}
  \end{figure}

Eq.~(\ref{sbs_3_wave})-(\ref{sbs_3_wave3}) are solved by Van Leer scheme (VL3), and the damping of three waves are neglected for simplicity. We consider three different cases: (1) without frequency modulation, (2) frequency modulation with $\eta = 1$, (3) frequency modulation with $\eta = 3$. As shown in Fig.~\ref{tr_ref}, the red line represents the time-dependence of SBS reflectivity without pump's frequency modulation, the average reflectivity is $3.42\%$, which agrees with the reflectivity obtained by Eq.~(\ref{gain}) showed by black dashed line. Blue line is the SBS reflectivity with pump's frequency modulation for $\eta = 1$, five bumps of reflectivity are observed, and the average reflectivity is $11.5\%$. These bumps come from the time window of negative velocity of resonant points, because the effective interaction length are much larger at negative velocity time window than that in positive velocity time window.

The green line in Fig.~\ref{tr_ref} indicates that the SBS reflectivity is significantly reduced when $\eta = 3$, which is even below the reflectivity without bandwidth case, and the average reflectivity of green line is $1\%$. The reason is that the effective interaction length decreases with bandwidth when $\eta \geq 1$. As shown in Fig.~\ref{ref_yita}, the SBS reflectivity first increase when $0\leq\eta\leq 0.8$ and then decreases when $\eta\geq 0.8$, which is agreed with the theoretical predictions above. The maximum reflectivity corresponds to $\eta = 0.8$, not $\eta = 1$ because the group velocity of seed laser is less than $c$ when it passes through the plasma.  We also observe that the average reflectivity is lower than that of no bandwidth case when $\eta \geq 2.3$, because when the bandwidth of pump laser is larger, the velocities of resonant points are much larger than the velocity of seed laser and the gain of seed laser will be less than $G_{r0}$. The effective interaction length can be approximated as $ L_{eff} = 2\sqrt{\frac{2\gamma_{0}^{2}}{|(|V_{1}|+|v_{r}|)V_{2}|}}$ when $\eta \geq 2.3$, because the negative velocity time window can be neglected.

The results of three-wave simulations show that in a flowing inhomogeneous plasma, the reflectivity of SBS is significantly increased when  the frequency modulation with $\eta \approx 1$ is added to pump laser, and the reflectivity is reduced  by larger-bandwidth pump with $\eta > 2.3$. Thus, in experiments, one should avoid the pump laser with $\eta \approx 1$ type frequency modulations, and choose the frequency-modulated pump laser with larger bandwidth to suppress SBS.

The evolution of SBS will enter into nonlinear region because of large amplitude of ion acoustic waves when the intensity of pump laser is high enough. The three-wave simulations did not consider the nonlinear effects of ion acoustic wave, such as nonlinear Landau damping, particle trapping and nonlinear frequency shift. In next section, we will use one-dimensional fully kinetic  Vlasov-Maxwell simulations to consider the effects of frequency modulation of pump laser in the nonlinear regime.

\section{Kinetic simulations with the frequency-modulated pump laser }\label{vlasov Simulation model}

\subsection{ One-dimensional Vlasov-Maxwell simulations }

\begin{figure}[htbp]
    \begin{center}
      \includegraphics[width=0.5\textwidth,clip,angle=0]{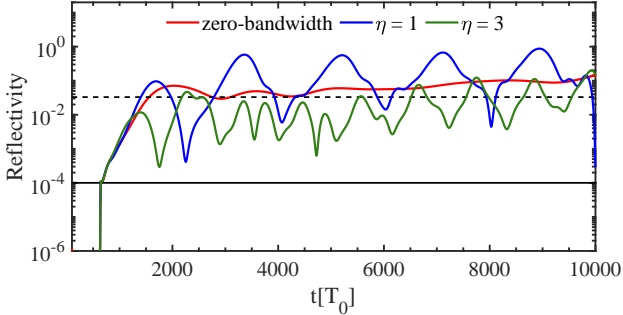}\vspace{-10pt}
      \caption{\label{vlasov_ref}  Vlama simulation results. The red line is the time dependence of SBS reflectivity with ordinary pump laser. The blue line is the time dependence of SBS reflectivity for $\eta = 1$ case. The green line is the time dependence of SBS reflectivity for $\eta = 3$ case. The black dashed line is obtained by the Rosenbluth theory.}
    \end{center}
  \end{figure}

The kinetic simulations included Vlasov-Maxwell simulations and PIC simulations are implemented to study the nonlinear effects of frequency modulations.

The one-dimensional Vlasov-Maxwell code called $Vlama$ uses VL3 scheme to solve Vlasov equation and uses fast Fourier transform to solve Possion equation.  In the $Vlama$ simulations,  the Helium plasma is used, where $Z =2$, $m_{i} = 7344 m_{e}$, and the temperature of electron and ion are $T_{e} = 1.5keV$ and $T_{i} = 0.5keV$, respectively. The ion acoustic velocity is obtained $C_{s} = 0.001152 c$. The flowing velocity of plasma is also $V(x) = C_{s}(7/4-3x/2x_{end})$. The simulation box is $L = 600 \lambda_{0}$, $10\%$ vacuum space is reserved in the both sides, so the plasma density ranges from $0.0667$ $n_{c}$ to $0.2$ $n_{c}$, and we choose the center point  $x_{0}$ to be the resonant point of SBS. The flowing velocity of plasma and plasma density are the same to that in three-wave simulations. The discrete spatial unit and temporal unit are $dx = 0.31416 c/\omega_{00}$ and $dt = 0.31416 \omega_{00}^{-1}$, respectively. The velocity spaces of electrons and ions in simulations are $[-0.8c,0.8c]$ and $[-0.01c,0.01c]$. The mesh grids of distribution functions of electrons and ions in the space-velocity plane are $12000\times2049$.

The wavelengths of unmodulated pump laser and seed laser are both $351\rm{nm}$, the intensities of pump laser and seed laser are same to that in three-wave simulations.
The pump laser is add a frequency modulation, $E_{y}(t) = a_{0}\cos[\omega_{00}t+\beta \sin(\omega_{m}t)]$, the bandwidth of pump laser is obtained by $\triangle\omega = 2\beta\omega_{m}$. Based on the analysis in Sec ~\ref{theoretical model}, when $\eta = 1$, $\triangle\omega = 1.7\times 10^{-3} \omega_{00} $, which is less than the bandwidth used in experiment $\triangle\omega = 3.0\times 10^{-3} \omega_{00}$\cite{bandwidth}. Thus, $\eta = 1$ is possible to achieve in current experiments, however the increase of SBS reflectivity by $\eta = 1$ type bandwidth laser has not been paid enough attention.

\begin{figure*}[htbp]
    \begin{center}
      \includegraphics[width=0.95\textwidth,clip,angle=0]{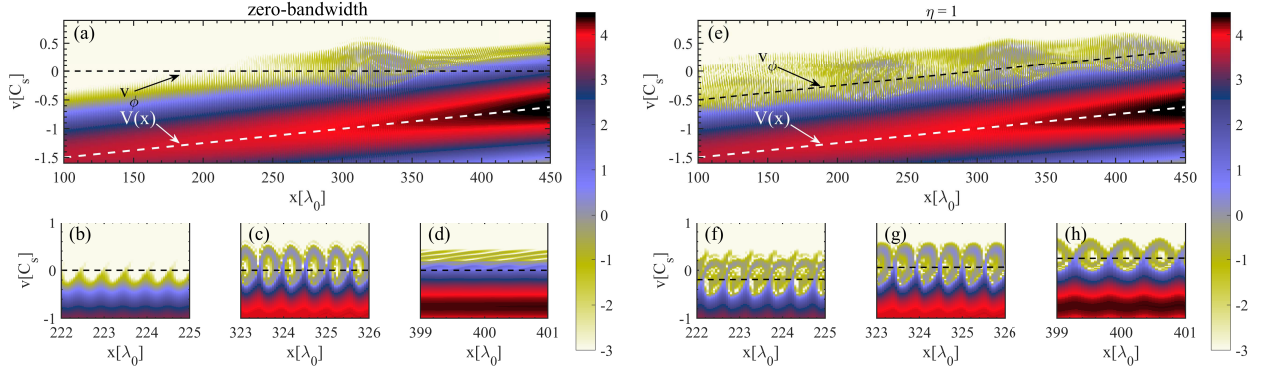}\vspace{-10pt}
      \caption{\label{ion_fig} Vlama simulation results. (a), (b), (c) and (d) are the particle trapping of ions in zero-bandwidth case. (e), (f), (g) and (h) are the particle trapping of ions in $\eta = 1$ case. Black dashed line represent the phase velocity of ion acoustic wave,and the white dashed line is the plasma flowing velocity.  }
    \end{center}
  \end{figure*}

  \begin{figure}[htbp]
    \begin{center}
      \includegraphics[width=0.49\textwidth,clip,angle=0]{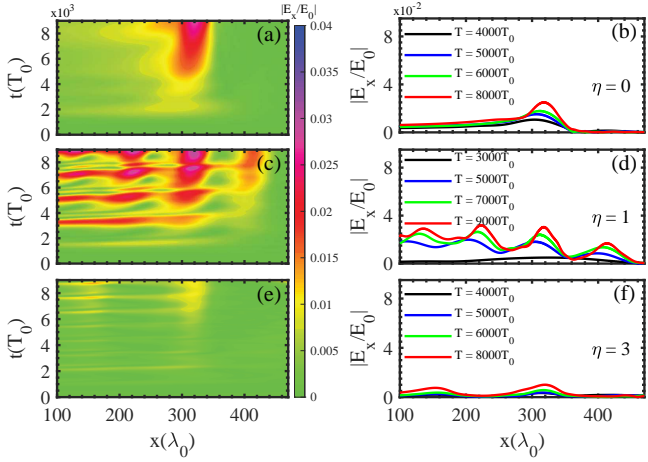}\vspace{-10pt}
      \caption{\label{vlasov_ex} (a) and (b) are the electrostatic amplitude evolution with time for zero-bandwidth case. (c) and (d) are the electrostatic amplitude evolution with time for $\eta =1$ case. (e) and (f) are the electrostatic amplitude evolution with time for $\eta =3$ case. }
    \end{center}
  \end{figure}

As shown in Fig.~\ref{vlasov_ref}, red line represents the time dependence of SBS reflectivity without frequency modulation by Vlama simulation, at early stage, $t < 4000T_{0}$, the evolution of SBS is agreed with the Rosenbluth theory (black dashed line), however, at $t > 4000T_{0} $, the reflectivity becomes larger than black dashed line, because when the amplitude of ion acoustic wave increases,   the frequency of ion acoustic wave decreases for the particle trapping effect, which will cancel with the detuning of non-uniform flowing velocity, the resonant point will move slowly to higher plasma density, this phenomenon is called autoresonance of SBS.  $\delta\omega_{i}$ is the nonlinear frequency shift because of particle trapping\cite{wangqing2}, \begin{equation}\label{nonlinear_shift}
         \frac{\delta\omega_{i}}{\omega_{20}} = -\zeta |\delta n_{i}/n_{i}|^{1/2},
\end{equation} $\omega_{20} = k_{2}C_{s}$ and $\zeta$ is the nonlinear frequency shift parameter defined as, \begin{equation}\label{shift_para}
  \zeta = \frac{1}{\sqrt{2\pi}}[\alpha_{i}\sqrt{\frac{ZT_{e}}{T_{i}}} (\upsilon^{4}-\upsilon^{2})e^{-\upsilon^{2}/2} - \alpha_{e}    ],
\end{equation}
where, $\alpha_{i} = 0.823$ is obtained by using sudden approximation for ions, and  $\alpha_{e} = 0.544$ is obtained by using adiabatic approximation for electrons, $\upsilon = (v_{\phi}-V(x))/v_{thi}$, $v_{\phi}$ is the phase velocity of ion acoustic waves and $v_{thi}$ is the thermal velocity of ions.

In Fig.~\ref{ion_fig} (a), the ion trapping is observed near $x_{0}$, the frequency will shift because of the formation of vortexes. The particle trapping happens when the amplitude of ion acoustic wave is large. In Fig.~\ref{vlasov_ex} (a) and (b), we can clearly observe that the peak of electrostatic field of ion acoustic wave locates near $x_{0}$ and moves towards to higher density in nonlinear region, which is the evidence of autoresonance. We observe that the phase velocity of ion acoustic wave in Fig.~\ref{ion_fig} (b), (c) and (d) are all equaled to $0$, which agrees earlier work\cite{wangqing2}, a consequence of the resonant excitation of pump laser and seed laser.

  \begin{figure*}
    \begin{minipage}[t]{0.487\linewidth}
    \centering
    \includegraphics[width=3.5in]{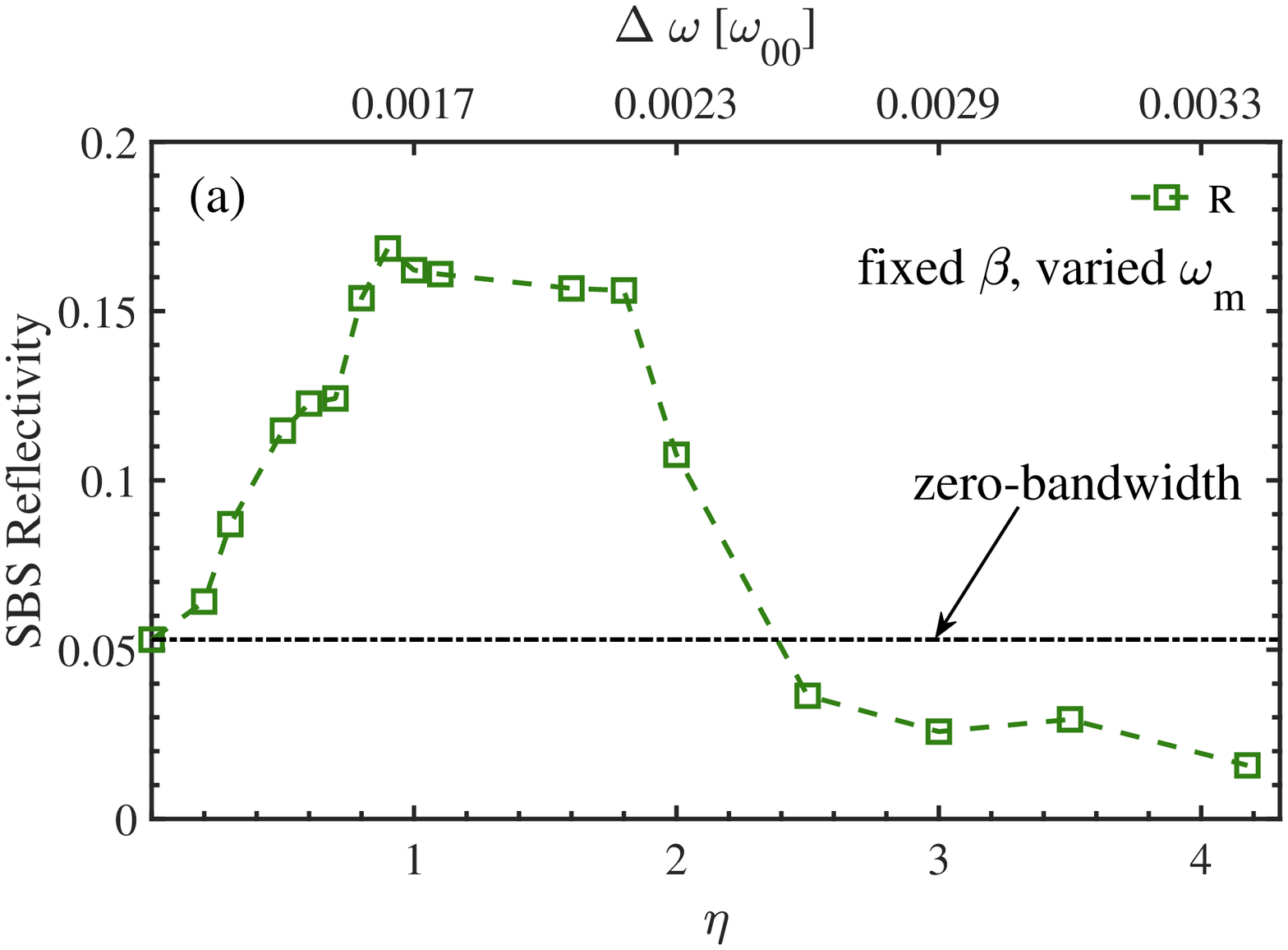}\vspace{-20pt}
    \end{minipage}%
    \begin{minipage}[t]{0.487\linewidth}
    \centering
    \includegraphics[width=3.5in]{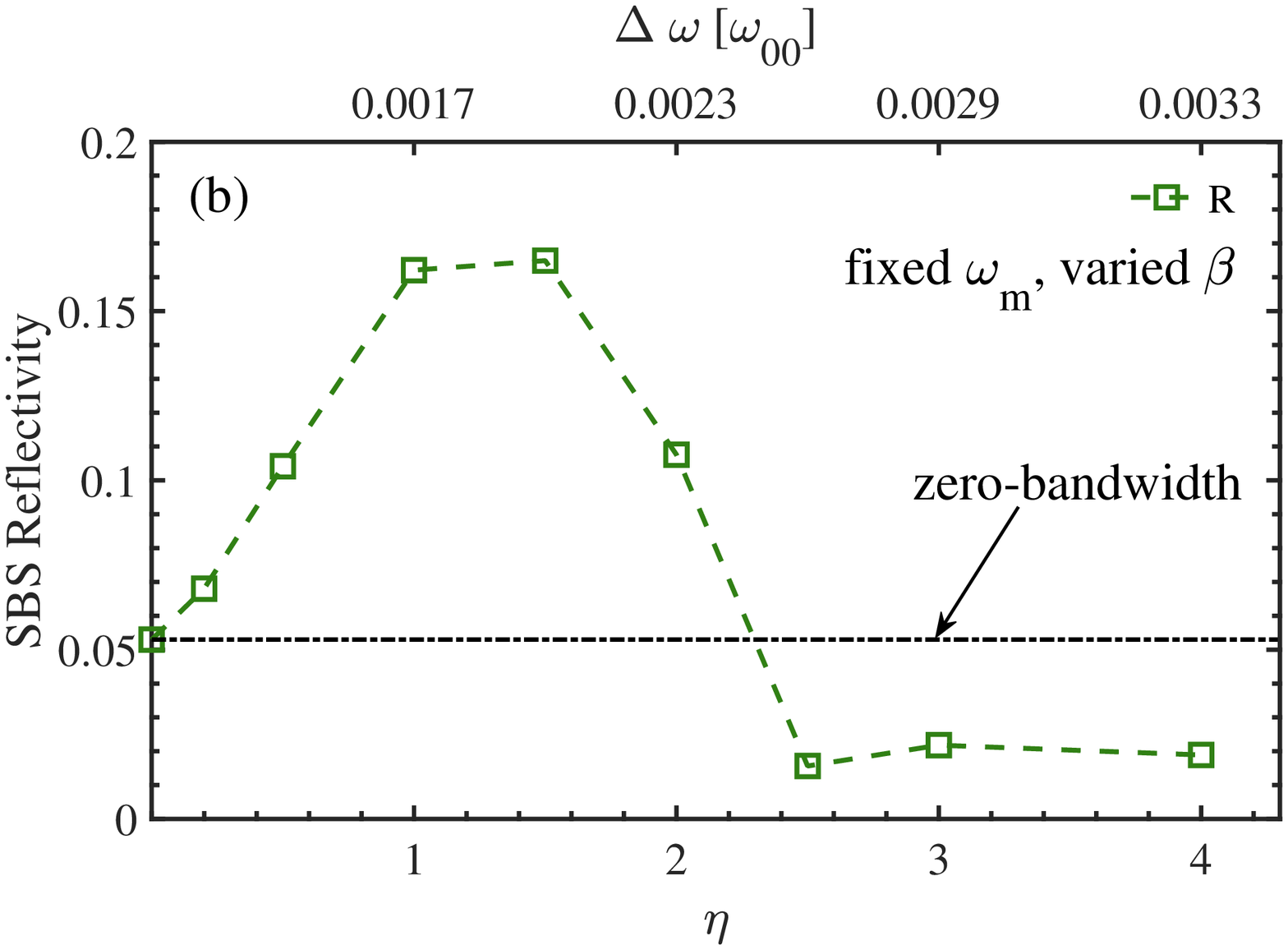}\vspace{-20pt}
    \end{minipage}

    \caption{\label{vlasov_yita} Vlama simulation results. (a)The dependence of bandwidth of pump on the reflectivity of SBS when $\beta$ is fixed. (b)The dependence of bandwidth of pump on the reflectivity of SBS when $\omega_{m}$ is fixed.  }
\end{figure*}

\begin{figure}[htbp]
    \begin{center}
      \includegraphics[width=0.5\textwidth,clip,angle=0]{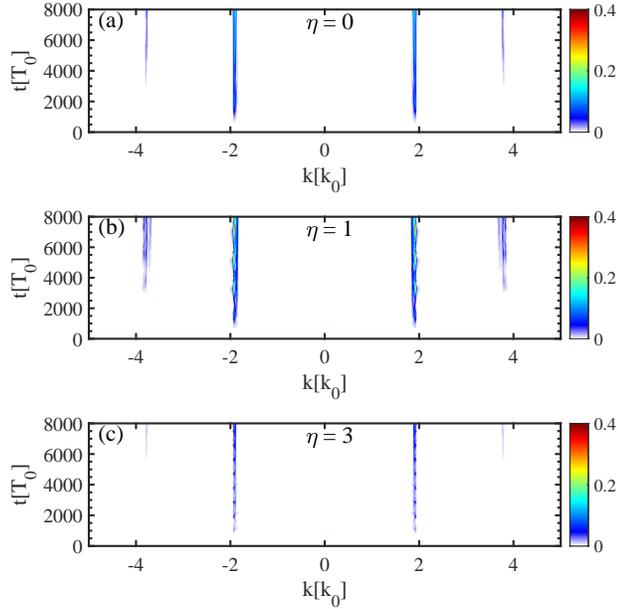}\vspace{-10pt}
      \caption{\label{iaw_kt} Three-wave simulation results. The green squares are the average SBS reflectivity for different frequency modulations. The black dashed line represents the average reflectivity without frequency modulation. The top axis is the corresponding bandwidth of pump laser for each $\eta$.   }
    \end{center}
  \end{figure}

The autoresonance is also found in $\eta = 1$ case, as shown in the blue line in Fig.~\ref{vlasov_ref}, there are five bumps, and the maximum points of bumps increases with time because of autoresonance. In Fig.~\ref{ion_fig} (e), the ion trapping at four different locations are observed. Because of the $\eta = 1$ type frequency modulation, the phase velocity of ion acoustic can be consider as a linear function of $x$, $V_{2}=V(x)+C_{s}$, which is agreed with the dispersion relation of ion acoustic wave in a flowing plasma. The multi-locations auto-resonantly of SBS results from the multi frequency of frequency-modulated pump laser, based on the Jacobi-Anger expansion, $E_{y}$ can be rewritten as
 \begin{equation}\label{anger_expan}
  E_{y}(t) = a_{0}Re[\sum^{+\infty}_{n=-\infty}J_{n}(\beta)e^{in\omega_{m}t+i\omega_{o}t}],
\end{equation} where Re(x) means the real part of the signal, $J_{n}(x)$ is the Bessel function of the first kind, Then, we apply the Fourier transform on $E_{y}(t)$,  obtain the spectrum of pump laser,
\begin{equation}\label{Ey_fft}
  |E_{y}(\omega)|^{2} = 2\pi^{2}a^{2}_{0}|\sum^{+\infty}_{n=-\infty}J_{n}(\beta)\delta(\omega\pm(\omega_{00}+n\omega_{m}))|^{2},
\end{equation} Where $\delta(x)$ is the Dirac delta function. We can know that the power spectrum of the frequency-modulated pump laser is comprised of multiple spectral lines, and the frequency difference between adjacent modes is $\omega_{m}$. Thus the frequency difference of iaws between two adjacent auto-resonant points is also $\omega_{m}$ if the frequency of seed laser is a constant.

 Correspondingly, the auto-resonantly growth of electrostatic field in Fig.~\ref{vlasov_ex} (c) and (d), there are four auto-resonant points, which is different from zero-bandwidth case.  When we add $\eta = 1$ type pump laser, the effective interaction length of SBS will increase, as shown in Fig.~\ref{vlasov_ex} (c), the interaction length in the negative velocity time window is about $300 \lambda_{0}$, the seed laser will be amplified when it passes through interaction region, the amplified seed laser will influence the SBS at lower plasma density. Then the ion acoustic wave at lower density will grow rapidly as shown in Fig.~\ref{vlasov_ex} (d). In another word, the frequency of seed does not change in the interaction region, and the change of phase velocity of ion acoustic wave results from the frequency modulation of pump laser. At lower density, the autoresonance also occurs when the frequency shift $\delta\omega_{i}$ cancels the wavenumber detuning term $iV_{2}\kappa'(x-x_{0})$. This kind of muti-locations autoresonance is first observed in this paper, which will also increase the reflectivity of SBS.

  \begin{figure}[htbp]
    \begin{center}
      \includegraphics[width=0.49\textwidth,clip,angle=0]{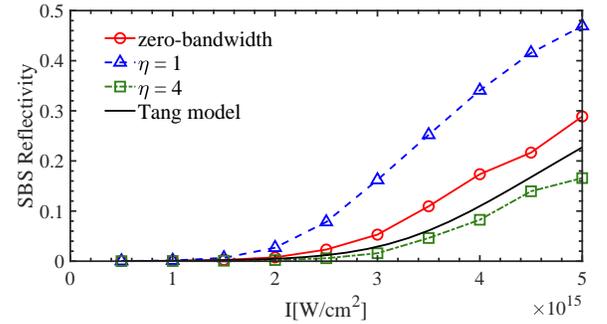}\vspace{-10pt}
      \caption{\label{ref_I} The change of SBS reflectivity for different pump intensities. Red circles are the reflectivity for zero-bandwidth case. Blue  triangles are the reflectivity for $\eta =1$ case. Green squares are the reflectivity for $\eta =4$ case. The black line is obtained by using Eq.~(\ref{tang}). }
    \end{center}
  \end{figure}

The green line in Fig.~\ref{vlasov_ref} shows the SBS reflectivity in $\eta = 3$ case, SBS is well suppressed at  $t < 4000 T_{0}$ because the reduction of effective interaction length. However, at the nonlinear stage $t > 4000T_{0}$, the autoresonance also happens, and the SBS reflectivity is larger than the Rosenbluth reflectivity (black dashed line). The autoresonance of SBS can also be observed in Fig.~\ref{vlasov_ex} (e) and (f).
 The average SBS reflectivity for different $\eta$ are shown in Fig.~\ref{vlasov_yita}, the average SBS reflectivity achieves maximum at $ \eta \approx 1.0$, and then decreases with $\eta$.  The average reflectivity is lower than no bandwidth case when $\eta > 2.3$. Similar results are observed either when $\beta$ is fixed or when $\omega_{m}$ is fixed.

 It is a known and popular fact that generation of ion sound harmonics draining the energy from the fundamental mode, which suppresses the SBS\cite{har,har2}, However, as shown in Fig.~\ref{iaw_kt} the wavnumber spectrum of IAW in the three different cases. the modes near $2k_{0}$ are the fundamental mode of IAW, and the modes near  $4k_{0}$ are the first harmonic waves of IAW. We find that the main difference occurs on the fundamental modes when the frequency-modulated pump laser is applied.

 In order to investigate the influence of frequency modulation on SBS with different intensities of pump laser, we change the intensity of pump laser from $5\times10^{14}W/cm^{2}$ to $5\times10^{15}W/cm^{2}$. As shown in Fig.~\ref{ref_I}, the black line represents the theoretical reflectivity by Tang's formula with considering pump depletion\cite{tang},\begin{equation}\label{tang}
  R = \frac{be^{(1-R)G_{r0}}}{1+b-R},
\end{equation}
where, $b = I_{seed0}/I_{pump}$ is the seed level, $R$ is the SBS reflectivity. The red circles in Fig.~\ref{ref_I} is the average SBS reflectivity without pump frequency modulation, it is larger than Tang model (black line) when $ I > 1.5\times 10^{15} W/cm^{2}$, because the autoresonance of SBS happens when the pump intensity is high enough to induce particle trapping. The blue triangles are the reflectivity of $\eta = 1$ cases, which is much larger than that of no bandwidth cases (red circles). Both longer effective interaction length and muti-locations autoresonance contribute to the high level reflectivity. The reflectivity of $\eta = 4$ cases are shown by green squares, we can observe that SBS is suppressed by the large-bandwidth frequency modulation, and the reflectivity is even less than the theoretical reflectivity (black line).



In conclusion,  $\eta \approx 1$ type frequency modulation of pump laser no only increase the effective interaction length but also lead to muti-locations autoresonance, the reflectivity is substantially increased. However, the large-bandwidth frequency modulation will reduce the effective interaction length and make it difficult for autoresonance to grow. Thus, large-bandwidth frequency modulation is an effective way to reduced the reflectivity of SBS.

\subsection{One-dimensional PIC simulations }

\begin{figure}[htbp]
    \begin{center}
      \includegraphics[width=0.47\textwidth,clip,angle=0]{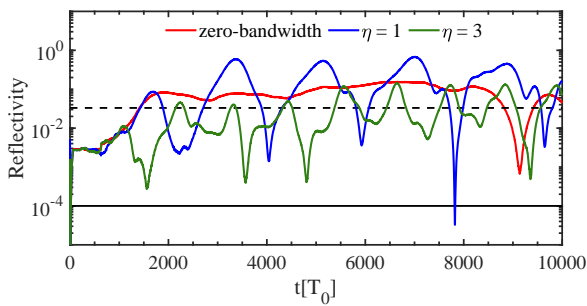}\vspace{-10pt}
      \caption{\label{pic_ref} PIC simulation results. The red line is the time dependence of SBS reflectivity with ordinary pump laser. The blue line is the time dependence of SBS reflectivity for $\eta = 1$ case. The green line is the time dependence of SBS reflectivity for $\eta = 3$ case. The black dashed line is obtained by the Rosenbluth theory.  }
    \end{center}
  \end{figure}

\begin{figure}[htbp]
    \begin{center}
      \includegraphics[width=0.47\textwidth,clip,angle=0]{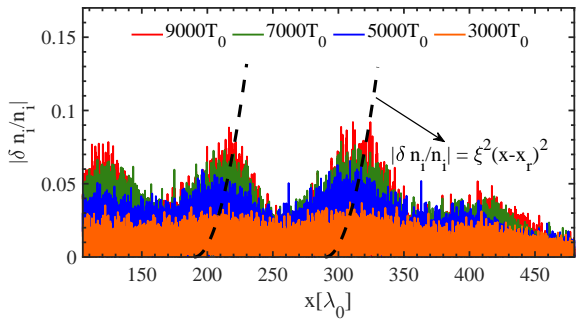}\vspace{-10pt}
      \caption{\label{delt_ne} PIC simulation results. The electron density perturbation at different times. We can observe four density peaks because of the autoresonance. The black dased lined is obtained by using  Eq.~(\ref{auto_re}).  }
    \end{center}
  \end{figure}

 PIC simulation is a well-known method for simulating laser plasma instability, which considers the kinetic and relativistic effects of particles. To verify the effects of frequency-modulated pump laser on SBS, we decide to use the one-dimensional PIC code EPOCH\cite{epoch}. The plasma parameters are same as the Vlasov simulations, and we consider zero-bandwidth case, $\eta = 1$ case, and $\eta = 3$ case.  The flowing velocity of plasma is also $V(x) = C_{s}(7/4-3x/2x_{end})$. There are $30$ cells for each wavelength of pump laser and $10^{4}$ particles in each cell for electrons and ions. We use large number of particles to reduce the thermal noise in PIC simulations. The intensity of pump laser is $3\times10^{15} W/cm^{2}$, and the seed level is $I_{seed0}/I_{pump} = 10^{-4}$. We note that the flowing velocity of plasma is set by the drifting velocity in PIC. $10\%$ vacuum space is reserved on both sides of the simulation box, and an open boundary condition is used for both waves and particles.

As shown in Fig.~\ref{pic_ref}, the red line is the reflectivity in no bandwidth case, it is larger than the Rosenbluth theory because of the autoresonance of SBS. When the $\eta = 1$ type frequency modulation is added in pump laser, there are also five bumps of reflectivity, which agree with the Vlasov simulations in Fig.~\ref{pic_ref}. When $\eta =3$, the SBS is suppressed below the Rosenbluth reflectivity at early stage, however the reflectivity increases because of the autoresonance. The results of PIC simulations agree with Vlama simulation results.

We also investigate the muti-location autoresonance in $\eta = 1$ case, $\upsilon = C_{s}/v_{thi}$ in Eq.~(\ref{shift_para}) is a constant in the interaction region, thus, the frequency shift parameter $\zeta$ is also a constant. As discussed in the early work, when the frequency shift cancels the wavenumber detuning, the resonant points moves by obeying\cite{wangqing2},
\begin{equation}\label{auto_re}
|\delta n_{i}/n_{i}| =\xi^{2}(x-x_{r})^{2}= (V_{2}\kappa'/\zeta k_{2}C_{s})^{2}(x-x_{r})^{2},
\end{equation}

As shown in Fig.~\ref{delt_ne}, the ion density perturbation $|\delta n_{i}/n_{i}|$ growths with time at $120 \lambda_{0}$, $200 \lambda_{0}$, $300 \lambda_{0}$ and $400 \lambda_{0}$. If there are autoresonance at these locations, the resonant points will move as time. The black dashed lines are plotted by using Eq.~(\ref{auto_re}), and we observe that the move of resonant points agree with the theoretical predication. Thus, we have demonstrated that autoresonance occurs at these locations because of the $\eta = 1$ type frequency modulation.

\section{Conclusion and discussion}\label{conclusion}

In this paper, first we construct a three-wave model to study the influence of frequency-modulated pump laser on SBS. We find that the reflectivity of SBS increases when $\eta$ is close to $1$, and the SBS will be suppressed when $\eta > 2.3$. The reason of suppression is that the velocities of resonant points are much larger than the group velocity of seed laser when $\eta > 2.3$, which reduces the effective interaction length. Then, the Vlama simulations are used to investigate the nonlinear effects induced by frequency modulations, we observe the muti-locations autoresonance when $\eta = 1$ type frequency modulation is applied. And the larger bandwidth modulation can also suppress the growth of SBS at nonlinear region. At last, the PIC simulations are used to prove our theoretical predications, and we prove that the growth of density perturbation are indeed induced by autoresonance, the move of resonant points agrees with theoretical predications.

In ICF experiments, pump laser with bandwidth is often used to mitigate the laser plasma instabilities. Early works show that the large bandwidth pump laser can restrain the growth of instabilities in homogeneous plasma. However, in a flowing inhomogeneous plasma, our results show that $\eta \approx 1$ type frequency modulation should be avoided, because it will increase the effective interaction length and even induce the  nuti-locations autoresonance. We believe that the larger bandwidth modulation is a promising way to reduce the reflectivity of SBS in experiments.

\section*{Acknowledgements}
We are pleased to acknowledge useful discussions with Q. Wang, Y. Z. Zhou, Y. G. Chen, N. Peng, S. Tan and W. B. Yao.  This work was supported by the Scientific Research Foundation for High-level Talents of Anhui University of Science and Technology (Grant No. 2022yjrc106), the Strategic Priority Research Program of Chinese Academy of Sciences (Grant No.XDA25050700), National Natural Science Foundation of China (Grant Nos.11975059),and Natural Science Foundation of Hunan Province, China (Grant No.2020JJ5029).

\begin{appendix}\label{my_appendix}
\section{The derivation of three-wave model with frequency-modulated pump}\label{AFluid theory}

The three-wave model of SBS in inhomogeneous plasmas starts with the coupling equations of SBS,
\begin{equation} \label{Atr_1}
[\partial_{t}^{2}-c^{2}\nabla^{2}+\omega_{pe}^{2}]A_{0}=-\frac{4\pi e^{2}}{m_{e}}
\tilde{n}_{e} A_{1},
\end{equation}
\begin{equation} \label{Atr_2}
[\frac{\partial^{2}}{\partial t^{2}}-c^{2}\nabla^{2}+\omega_{pe}^{2}]A_{1}=-\frac{4\pi e^{2}}{m_{e}}
\tilde{n}_{e} A_{0},
\end{equation}

\begin{equation} \label{Atr_3}
    [(\partial_{t}+V\cdot\nabla)^{2}-C_{s}^{2}\nabla^{2}]\tilde{n}_{e}=\frac{Zn_{0}e^{2}}{m_{e}m_{i}c^{2}}\nabla^{2}(A_{0}\cdot A_{1}),
\end{equation}where, $A_{0}$, $A_{1}$ are the vector potential for the pump and seed lasers, $\tilde{n}_{e}$ electron density perturbation associated with IAW, since quasi-neutrality requires $\tilde{n}_{e} = Z\tilde{n}_{i}$. $V$ is the plasma flowing velocity.

As we know, a laser with a small bandwidth $\Delta\omega$ has the coherence time $\tau \approx 2\pi/\Delta\omega$. In this paper, the bandwidth of pump is in the level of $10^{-3}\omega_{00}$, so $\tau$ is on the level of $10^{3}T_{0}$, where $T_{0}$ is the single period of pump laser, then the coherence length of pump laser is obtained by $L_{cohe} = c\tau$, which is on the level of $10^{3}\lambda_{0}$. The coherence length of pump laser is much larger than the wavelength of pump laser, $i.e.$ $L_{cohe} \gg \lambda_{0}$, so the pump lasers with small bandwidth still have $|\partial_{t}A_{0}|\ll|\omega_{00}A_{0}|$ and $|\bigtriangledown A_{0}|\ll|k_{0}A_{0}|$. We could consider the bandwidth effects in a much larger spatiotemporal scale. Also, the coherence length of pump laser is larger than the length of plasma region. Thus the frequency-modulated pump laser in our paper can be expressed as the product of a slowly changing amplitude envelope and a rapidly oscillating periodic function,

\begin{equation} \label{Atr_4}
\begin{split}
&A_{0} = \frac{1}{2}\tilde{A}_{0}e^{-i\omega_{00}t+i\int_{0}^{t}\delta \omega(x,t') dt'+i\int_{0}^{x}k_{0}(x')\cdot x'} + c.c.\\
&A_{1} = \frac{1}{2}\tilde{A}_{1}e^{-i\omega_{1}t+i\int_{0}^{x}k_{1}(x')\cdot x'} + c.c.\\
&\tilde{ n}_{e} = \frac{1}{2}\delta n_{e}e^{-i\omega_{2}t+i\int_{0}^{x}k_{2}(x')\cdot x'} + c.c.
\end{split}
\end{equation} where $\delta\omega(x,t)=\omega_{m}\beta cos(\omega_{m}t-\frac{\omega_{m}}{c}x)$ is the frequency change of pump laser. In recent work, the pump laser with small bandwidth is also approximated by envelope form to study the threshold of laser plasma instabilities\cite{lpse}. Substituting Eq.~(\ref{Atr_4}) to the coupling equations, then we retain the first-order terms and neglect the second-order terms.   the envelope form of coupling equations is obtained,
\begin{equation} \label{Atr_5}
(\partial_{t}+V_{0}\nabla+\nu_{0})\tilde{A}_{0}=- \frac{i\pi e^{2}}{m_{e}\omega_{00}}\delta n_{e}\tilde{A}_{1}e^{-i\varphi},
\end{equation}

\begin{equation} \label{Atr_6}
(\partial_{t}+V_{1}\nabla+\nu_{1})\tilde{A}_{1}=- \frac{i\pi e^{2}}{m_{e}\omega_{1}}\delta n_{e}^{*}\tilde{A}_{0}e^{i\varphi},
\end{equation}

\begin{equation} \label{Atr_7}
(\partial_{t}+V_{2}\nabla+\nu_{2})\delta n_{e}=- \frac{in_{0}Zk_{2}^{2}e^{2}}{4k_{2}C_{s}m_{i}m_{e}c^{2}}\tilde{A}_{0}\tilde{A}_{1}^{*}e^{i\varphi},
\end{equation}
where $\varphi = \int_{0}^{t}\delta \omega(x,t') dt'+\int_{0}^{x}dx'[k_{0}(x')-k_{1}(x')-k_{2}(x')] \approx \int_{0}^{t}\delta \omega(x,t') dt'+\int_{0}^{x}dx'[\kappa'(x'-x_{0})]$.

 Let's assume that $\varphi_{1} = \int_{0}^{t}\delta \omega(t') dt'$ and $\varphi_{2} = \int_{0}^{x}dx'[\kappa'(x'-x_{0})]$, which leads to $\varphi = \varphi_{1}+\varphi_{2}$, then the term $ e^{-i\varphi_{2}}$ can be absorbed in $\delta n_{e}$, $\delta n_{e} e^{-i\varphi_{2}}  \rightarrow \delta n_{e}$, we get,
 \begin{equation} \label{Atr_11}
(\partial_{t}+V_{0}\nabla+\nu_{0})\tilde{A}_{0}=- \frac{i\pi e^{2}}{m_{e}\omega_{00}}\delta n_{e}\tilde{A}_{1}e^{-i\varphi_{1}},
\end{equation}

\begin{equation} \label{Atr_12}
(\partial_{t}+V_{1}\nabla+\nu_{1})\tilde{A}_{1}=- \frac{i\pi e^{2}}{m_{e}\omega_{1}}\delta n_{e}^{*}\tilde{A}_{0}e^{i\varphi_{1}},
\end{equation}

\begin{equation} \label{Atr_13}
(\partial_{t}+V_{2}\nabla+\nu_{2}+iV_{2}\kappa'(x-x_{0}))\delta n_{e}=- \frac{in_{0}Zk_{2}^{2}e^{2}}{4k_{2}C_{s}m_{i}m_{e}c^{2}}\tilde{A}_{0}\tilde{A}_{1}^{*}e^{i\varphi_{1}},
\end{equation} we substitute $a = eA/m_{e}c$  to coupling equations, the space is normalized to $c\omega_{00}$, time is normalized to $\omega_{00}^{-1}$, and $\delta n_{e}$ is normalized to the critical density of pump laser, $n_{c}$, We finally get the three-wave model of SBS with frequency-modulated pump,
\begin{equation}\label{Asbs_3_wave}
    (\partial_{t}+V_{0}\partial_{x}+\nu_{0})a_{0}=-\frac{i}{4}Z\delta n_{i}a_{1}e^{-i\varphi_{1}},
\end{equation}

\begin{equation}\label{Asbs_3_wave2}
    (\partial_{t}+V_{1}\partial_{x}+\nu_{1})a_{1}=-\frac{i\omega_{00}}{4\omega_{1}}Z\delta n_{i}^{\ast}a_{0}e^{i\varphi_{1}},
\end{equation}

\begin{equation}\label{Asbs_3_wave3}
    (\partial_{t}+V_{2}\partial_{x}+\nu_{2}+iV_{2}\kappa'(x-x_{0}))\delta n_{i}=-\frac{-i4\gamma_{0}^{2}c^{2}}{Z\omega_{00}^{2}v_{osc}^{2}}a_{0}a_{1}^{\ast}e^{i\varphi_{1}},
\end{equation}

\end{appendix}

\begin{thebibliography}{100}
\newcommand{\DOI}[1]{doi: \href{https://doi.org/#1}{#1}}

\bibitem{ICF1}R. Betti, C. D. Zhou, K. S. Anderson, L. J. Perkins, W. Theobald, and A. A. Solodov,  Phys. Rev. Lett. {\bf 98}, 155001 (2007).(\DOI{10.1103/PhysRevLett.98.155001})
 \bibitem{ICFn}V. Tikhonchuk, Y. J. Gu, O. Klimo, J. Limpouch, and S. Weber, Matter Radiat. Extremes {\bf 4}, 045402 (2019).(\DOI{10.1063/1.5090965})
\bibitem{ICF2}Basov N. G., Guskov S. Y. and Feokistov L. P., J. Russ. Laser Res. {\bf 13}, 396 (1992).(\DOI{10.1007/BF01124892})
\bibitem{ICF3}Max Tabak, James Hammer, Michael E. Glinsky, William L. Kruer, Scott C. Wilks, John Woodworth, E. Michael Campbell, and Michael D. Perry,  Phys. Plasmas {\bf 1}, 1626 (1994).(\DOI{10.1063/1.870664})
\bibitem{D_D}  S. E. Bodner, D. G. Colombant, J. H. Gardner, R. H. Lehmberg, S. P. Obenschain, L. Phillips, A. J. Schmitt, J. D. Sethian, R. L. McCrory, W. Seka, C. P. Verdon, J. P. Knauer, B. B. Afeyan, and H. T. Powell, Phys. Plasmas {\bf 5}, 1901 (1998).(\DOI{10.1063/1.872861})
\bibitem{I_D}   J. Lindl, Phys. Plasmas {\bf 2}, 3933 (1995).(\DOI{10.1063/1.871025})
\bibitem{H_D} X. T. He, J.W. Li, Z. F. Fan, L. F.Wang, J. Liu, K. Lan, J. F.Wu, and W. H. Ye, Phys. Plasmas {\bf 23}, 082706 (2016).(\DOI{10.1063/1.4960973})
\bibitem{S_I} X. Ribeyre, G. Schurtz, M. Lafon, S. Galera, and S. Weber, Plasma Phys. Control. Fusion {\bf 51}, 015013 (2009).(\DOI{10.1088/0741-3335/51/1/015013})
\bibitem{sbs1} H. A. Rose and D. F. DuBois, Phys. Rev. Lett. {\bf 72}, 2883 (1994).(\DOI{10.1103/PhysRevLett.72.2883})

\bibitem{sbs2} P. Neumayer, R. L. Berger, L. Divol, D. H. Froula, R. A. London, B. J. MacGowan, N. B. Meezan, J. S. Ross, C. Sorce, L. J. Suter, and S. H. Glenzer, Phys. Rev. Lett. {\bf 100}, 105001 (2008).(\DOI{10.1103/PhysRevLett.100.105001})

\bibitem{sbs3} J. Li, S. Zhang, C. M. Krauland, H. Wen, F. N. Beg, C. Ren, and M. S. Wei, Phys. Rev. E {\bf 101}, 033206 (2020).(\DOI{10.1103/PhysRevE.101.033206})

\bibitem{sbs4} S. Zhang, J. Li, C. M. Krauland, F. N. Beg, S. Muller, W. Theobald, J. Palastro, T. Filkins, D. Turnbull, D. Haberberger, C. Ren, R. Betti, C. Stoeckl, E. M. Campbell, J. Trela, D. Batani, R. H. H. Scott, and M. S. Wei, Phys. Rev. E {\bf 103}, 063208 (2021).(\DOI{10.1103/PhysRevE.103.063208})

\bibitem{sbs5} C. Z. Xiao, Y. G. Chen, J. F. Myatt, Q. Wang, Y. Chen, Z. J. Liu, C. Y. Zheng, and X. T. He, Phys. Rev. E {\bf 104}, 065203 (2021).(\DOI{10.1103/PhysRevE.104.065203})

\bibitem{kruer}W. L. Kruer, \emph{The Physics of Laser Plasma Interactions} (Westview Press, Boulder, 2003).
\bibitem{Nicholson} D. R. Nicholson, \emph{Introduction to Plasma Theory }(John Wiley \& Sons, New York, 1983).

\bibitem{cpps} Z. J. Liu, B. Li, X. Y. Hu, J. Xiang, C. Y. Zheng, L. H. Cao, and L. Hao, Phys. Plasmas {\bf 23}, 022705 (2016).(\DOI{10.1063/1.4941967})

\bibitem{ssd} S. Skupsky, R. W. Short, T. Kessler, R. S. Craxton, S. Letzring, and J. M. Soures, J. Appl. Phys. {\bf 66}, 3456 (1989).(\DOI{10.1063/1.344101})

\bibitem{stud} B. J. Albright, L. Yin, and B. Afeyan, Phys. Rev. Lett. {\bf 113}, 045002 (2014).(\DOI{10.1103/PhysRevLett.113.045002})

\bibitem{liu_ps1} Z. J. Liu, C. Y. Zheng, L. H. Cao, B. Li, J. Xiang, and L. Hao, Phys. Plasmas 24, 032701 (2017).(\DOI{10.1063/1.4977910})

\bibitem{liu_ps} Z. J. Liu, Q. Wang, W. S. Zhang, B. Li, P. Li, W. G. Zheng, X. Li, J. W. Li, L. H. Cao, Y. K. Ding, and X. T. He, Phys. Plasmas {\bf 30}, 032703 (2023).(\DOI{10.1063/5.0137403})

\bibitem{qing} Q. K. Liu, E. H. Zhang, W. S. Zhang, H. B. Cai, Y. Q. Gao, Q. Wang, and S. P. Zhu, Phys. Plasmas  {\bf 29}, 102105 (2022).(\DOI{10.1063/5.0105089})

\bibitem{luo_1} M. F. Luo, S. H\"{u}ller, M. Chen, and Z. M. Sheng, Phys. Plasmas {\bf 29}, 032102 (2022).(\DOI{10.1063/5.0078985})
\bibitem{luo_2} M. F. Luo, S. H\"{u}ller, M. Chen, and Z. M. Sheng, Phys. Plasmas {\bf 29}, 072709 (2022).(\DOI{10.1063/5.0096771})

\bibitem{zhao_1} Y. Zhao, L. L. Yu, J. Zheng, S. M. Weng, C. Ren, C. S. Liu, and Z. M.
Sheng, Phys. Plasmas {\bf 22}, 052119 (2015).(\DOI{10.1063/1.4921659})

\bibitem{zhao_2} Y. Zhao, S. M. Weng, M. Chen, J. Zheng, H. B. Zhuo, C. Ren, Z. M. Sheng, and J. Zhang, Phys. Plasmas {\bf 24}, 112102 (2017).(\DOI{10.1063/1.5003420})

\bibitem{VI1} V.I. Arkhipenko, V.N. Budnikov,  E.Z. Gusakov, N.M. Kaganskaya , V.L. Selenin  and L.V. Simonchik, Plasma Phys. Control. Fusion {\bf 37}, 1353 (1995).

\bibitem{VI2} V. I. Arkhipenko, E.Z. Gusakov, L. V. Simonchik, F. M. Truhachev, Phys. Rev. Lett.,{\bf 101}, 175004 (2008).
\bibitem{VI3} V. I. Arkhipenko, E.Z. Gusakov, V. A. Pisarev, L. V. Simonchik and B.O. Yakovlev, Phys. Plasmas {\bf 1}, 71 (2004).

 \bibitem{ma} H. H. Ma, X. F. Li, S. M. Weng, Z. M. Sheng, and J. Zhang, Matter Radiat. Extremes {\bf 6}, 055902 (2021).(\DOI{10.1063/5.0054653})
\bibitem{cbet} J. W. Bates, J. F. Myatt, J. G. Shaw, R. K. Follett, J. L. Weaver, R. H. Lehmberg, and S. P. Obenschain, Phys. Rev. E {\bf 97}, 061202(R) (2018).(\DOI{10.1103/PhysRevE.97.061202})
\bibitem{wen} H. Wen, R. K. Follett, A. V. Maximov, D. H. Froula, F. S. Tsung, and J. P. Palastro, Phys. Plasmas {\bf 28}, 042109 (2021).(\DOI{10.1063/5.0036768})

\bibitem{HaoL1} L. Hao, J. Li, W. D. Liu, R. Yan, and C. Ren, Phys. Plasmas {\bf 23}, 042702 (2016). (\DOI{10.1063/1.4945647})
\bibitem{Klimo1} O. Klimo, S. Weber, V. T. Tikhonchuk and J. Limpouch, Plasma Phys. Control. Fusion {\bf 52},  055013 (2010).  (\DOI{10.1088/0741-3335/52/5/055013})
\bibitem{Klimo2}  O. Klimo, V. T. Tikhonchuk, X. Ribeyre, G. Schurtz, C. Riconda, S. Weber, and J. Limpouch, Phys. Plasmas {\bf 18}, 082709 (2011).(\DOI{10.1063/1.3625264})
\bibitem{wangqing2} Q. Wang, C. Y. Zheng, Z. J. Liu, L. H. Cao, Q. S. Feng, C. S. Liu and X. T. He, Plasma Phys. Control. Fusion {\bf 61}, 085017 (2019).(\DOI{10.1088/1361-6587/ab2736})

\bibitem{wangqing1}  Q. Wang, C. Y. Zheng, Z. J. Liu, C. Z. Xiao, Q. S. Feng, H. C. Zhang and X. T. He, Plasma Phys. Control. Fusion {\bf 60}, 025016 (2018).(\DOI{10.1088/1361-6587/aa98bb})

\bibitem{rosenbluth1} M. N. Rosenbluth, Phys. Rev. Lett. {\bf 29}, 565 (1972). (\DOI{10.1103/PhysRevLett.29.565})
\bibitem{rosenbluth2} M. N. Rosenbluth, R. B. White, C. S. Liu, Phys. Rev. Lett. {\bf 31}, 1190 (1973).(\DOI{10.1103/PhysRevLett.31.1190})
\bibitem{bandwidth} S. Obenschain, N. Luhmann Jr, and P. Greiling, Phys. Rev. Lett. {\bf 36}, 1309 (1976). (\DOI{10.1103/PhysRevLett.36.1309})

\bibitem{har} L. M. Gorbunov, Sov. Phys. JETP {\bf 38}, 490 (1974).
 \bibitem{har2}   S. H\"{u}ller, Phys. Fluids B {\bf 3}, 3317 (1991).

\bibitem{tang} C. L. Tang, J. Appl. Phys., {\bf 37}, 2945 (1966).(\DOI{10.1063/1.1703144})

\bibitem{epoch} T. D. Arber, K. Bennett, C. S. Brady, A. Lawrence-Douglas, M. G. Ramsay, N. J. Sircombe, P. Gillies, R. G. Evans, H. Schmitz, A. R. Bell and C. P. Ridgers, Plasma Phys. Control. Fusion {\bf 57}, 113001 (2015). (\DOI{10.1088/0741-3335/57/11/113001})

\bibitem{lpse} R. K. Follett, J. G. Shaw, J. F. Myatt, C. Dorrer, D. H. Froula, and J. P. Palastro, Phys. Plasmas, {\bf 26}, 062111 (2019).(\DOI{10.1063/1.5098479})

\end{thebibliography}

\end{document}